\documentclass[pra,twocolumn,floatfix,superscriptaddress]{revtex4}

\usepackage{times}
\usepackage{graphicx}

\begin{document} 

\title{Linking Entanglement and Quantum Phase Transitions via Density Functional Theory}
\author{L.-A. Wu}
\affiliation{Chemical Physics Theory Group, Department of Chemistry, and Center
for Quantum Information and Quantum Control, University of Toronto, 80 St.
George St., Toronto, Ontario, M5S 3H6, Canada}
\author{M. S. Sarandy}
\affiliation{Escola de Engenharia Industrial Metal\'urgica de Volta Redonda, 
Universidade Federal Fluminense, Av. dos Trabalhadores 420, Volta Redonda,  
27255-125, RJ, Brazil.}
\affiliation{Instituto de F\'{\i}sica de S\~ao Carlos, 
Universidade de S\~ao Paulo, S\~ao Carlos, 13560-970, Brazil.} 
\author{D. A. Lidar}
\affiliation{Department of Chemistry and Department of 
Electrical Engineering-Systems, University of Southern California, Los Angeles, CA 90089}
\author{L. J. Sham}
\affiliation{Department of Physics, University of California San Diego, 9500
Gilman Drive, La Jolla, CA 92093}

\begin{abstract}
Density functional theory (DFT) is shown to provide a novel conceptual and
computational framework for entanglement in interacting many-body quantum
systems. DFT can, in particular, shed light on the intriguing relationship
between quantum phase transitions and entanglement. We use DFT concepts to
express entanglement measures in terms of the first or second derivative of
the ground state energy. We illustrate the versatility of the DFT approach
via a variety of analytically solvable models. As a further application we
discuss entanglement and quantum phase transitions in the case of mean field
approximations for realistic models of many-body systems.
\end{abstract}

\maketitle

\section{Introduction}

Density functional theory (DFT)~\cite{HK:64,Kohn:65} is to date the most
successful method for first principles calculations of the electronic
properties of solids. The key to its success is a transformation of the
dependence of the properties of a system of interacting particles on their
single particle potential, to a dependence on the ground state density,
thereby facilitating useful approximations of the many-body interaction for
first principles computations. One rather relevant phenomenon in many-body
physics is the ocurrence of quantum phase transitions (QPTs), which consist
in critical changes in the properties of the ground state, driven purely by
quantum fluctuations and effectively occurring at temperature $T=0$~\cite%
{Sachdev:book}. QPTs are associated with level crossings, which usually lead
to the presence of non-analyticities in the energy spectrum. Specifically, a
first-order QPT (1QPT) is characterized by a finite discontinuity in the
first derivative of the ground state energy. Similarly, a second-order QPT
(2QPT) is characterized either by a finite discontinuity or divergence in
the second derivative of the ground state energy, assuming the first
derivative is continuous.

Many-body physics and, in particular, critical phenomena near QPTs, have
recently been the subject of intense interest from the perspective of the
theory of quantum information. A key observation is that, since entanglement
describes correlations in a quantum system, its quantification may provide a
convenient and precise description of a QPT. Indeed, entanglement has been
found to exhibit scaling behavior near a critical point~\cite%
{Osterloh:02,Osborne:02,Vidal:03}. Moreover, under well-delineated conditions
and for distinguishable systems up to two-body interactions, a formal
relationship between QPT and bipartite entanglement was recently established~%
\cite{WuSarandyLidar:04}. Here, we show that entanglement may be well
specified and conveniently computed within DFT. In DFT, any entanglement
measure is a function(al) of the expectation values of the observables. This
procedure introduces a direct connection between entanglement and the
derivatives of the ground state energy of the quantum system with respect to
the field coefficients, leading to a deep relationship between entanglement
and QPT.

\section{Generalized Hohenberg-Kohn (HK) theorem and entanglement}
Consider a quantum system described by a Hamiltonian composed of two parts
\begin{equation}
H=H_{0}+H_{\mathrm{ext}}=H_{0}+\sum_{l}\lambda _{l}\widehat{A}_{l},
\label{eq00}
\end{equation}%
where $\lambda _{l}$ is the \textquotedblleft field
coefficient\textquotedblright\ (control parameter) associated with a set of
Hermitian operators $\{\widehat{A}_{l}\}$, e.g., an
observable relevant to driving a quantum phase transition. The index $l$ can
be discrete or continuous. The expectation values of $\widehat{A}_{l}$ for
a ground state $\left\vert \psi \right\rangle $ are denoted by the set $%
\{a_{l}\}\equiv \{\left\langle \psi \right\vert \widehat{A}_{l}\left\vert
\psi \right\rangle \}$.

DFT is originally based on the Hohenberg-Kohn (HK) theorem~\cite{HK:64}. In
the case of a many-electron system, the HK theorem establishes that the
ground state electronic density $n(\mathbf{r})$, instead of the potential $v(%
\mathbf{r})$, can be used as the fundamental variable to describe the
physical properties of the system. In the case of a Hamiltonian given by
Eq.~(\ref{eq00}), the HK theorem can be generalized to the statement that
there is a duality (in the sense of a Legendre transform) between the set of
expectation values $\{a_{l}\}$ (corresponding to $n(\mathbf{r})$) and the
set of field parameters $\{\lambda _{l}\}$ (corresponding to $v(\mathbf{r})$%
) ~\cite{Schonhammer:95}.
The commutativity of the densities at distinct points,
$[\widehat{n}(\mathbf{r}),\widehat{n}(\mathbf{r}^{\prime })]=0$ for $%
\mathbf{r\neq r}^{\prime }$, is a property of the original HK theorem.
%%%%%%%%%%%%%%%%%%%%%%
%%%% Adjusted in v6b:
%%In a lattice system, we will require that operators $A_l$ 
%%acting on different sites constitute a set of 
%%mutually commuting operators, so that $\{a_{l}\}$ 
%%may be regarded as a set of 
%%simultaneously 
%%measurable properties. 
In a lattice system, we require that the physical observables $\{A_l\}$ on different sites 
are mutually commuting operators. 
%%%%%%%%
This allows not only 
different observables on the same site, e.g. $S^x_l, S^y_l$, to be 
non-commutative, but also, for later use, endows a function of 
observables on different sites with a single site locality, such as 
the set of two site operators, $\{A_lA_{l+c}\}$, $l$ ranging over all 
sites and $c$ being a constant.
It follows from the Legendre transform that the ground state 
expectation value of any
observable can be interchangeably viewed as a unique function of either $%
\{\lambda _{l}\}$ or $\{a_{l}\}$.
(See Appendix~\ref{a1} for a simple proof of the HK theorem in a lattice).
Such a general duality has allowed for the
application of DFT in, e.g., interacting quantum spin systems~\cite%
{Libero:03}. Moreover, as we show below, it can provide a natural connection
between entanglement and QPT. Indeed, using the Hellmann-Feynman theorem~%
\cite{Hellmann:37,Feynman:39},
\begin{equation}
\frac{\partial E}{\partial \lambda _{l}}=\left\langle \psi \right\vert \frac{%
\partial H}{\partial \lambda _{l}}\left\vert \psi \right\rangle
=\left\langle \psi \right\vert \widehat{A}_{l}\left\vert \psi \right\rangle
=a_{l}.  \label{eq01}
\end{equation}%
This means that the set of observables $\{{\partial H}/{\partial \lambda _{l}%
}\}$ has a direct linear relation with $\{a_{l}\}$.
An example is the metallization of a semiconductor under pressure to a
value at which the band gap given by the discontinuity of the density
functional derivative of the ground state energy goes to zero~\cite%
{Sham:83,Perdew:83}.

\emph{The HK theorem can be used to redefine entanglement measures in terms
of new physical quantities: expectation values of observables, }$\{a_{l}\}$%
\emph{, instead of external control parameters}, $\{\lambda _{l}\}$.
Consider an arbitrary entanglement measure $M$ for the ground state of
Hamiltonian~(\ref{eq00}). We will focus here on bipartite entanglement, but
our discussion applies equally well to multipartite measures. We then prove
a central lemma, which very generally connects $M$ and energy derivatives.

\noindent\textbf{Lemma.} Any entanglement measure $M$ can be expressed as a
unique functional of the set of first derivatives of the ground state
energy:
\begin{equation}
M=M(\{a_{l}\})=M(\{\frac{\partial E}{\partial \lambda _{l}}\}),  \label{eq02}
\end{equation}
%%where the set of expectation values $\{a_{l}\}$ is taken over an 
%%arbitrary ground state wave function.
%%RESET
assuming that the ground state is non-degenerate.

\noindent \textbf{Proof.} Intuitively, the proof follows from the fact that,
according to the generalized HK theorem, any ground state wave function
$|\Psi \rangle$ is a unique functional of $\{a_{l}\}$
%%%%%%%%%%%%%%%%%%%
%% Adjusted in v6b:
%%RESET (in the case of a degenerate ground state, we will have a set of 
%%independent ground state wave functions). 
%%Since the ground state wave functions
%%%%%%%%%%%%%%%%%%%%%%% 
and since $|\Psi \rangle$ provides a complete 
description of the state of
the system, everything else is a unique functional of $\{a_{l}\}$ as 
well, including
$M$. More formally, let us consider the case of pairwise entanglement of
qubits. The case of higher dimensional systems or multipartite entanglement
is a direct generalization. Then: (A) $M_{ij}$ (entanglement measure between
qubits $i$ and $j$) is always a function $f$ of the matrix elements of the
2-qubit reduced density matrix $\rho _{ij}$: $M_{ij}=f(\rho _{ij})$. (B) The
matrix elements $\rho _{ij}$ are combinations of correlation functions $%
\langle \sigma _{i}^{a}\sigma _{j}^{b}\rangle =\mathrm{Tr}(\sigma
_{i}^{a}\sigma _{j}^{b}\rho _{ij})$, where $a,b=0,..,3$, with $\sigma _{0}=I$
(identity). This follows from an expansion of $\rho _{ij}$ in the Pauli
basis $\{\sigma _{i}^{a}\sigma _{j}^{b}\}$. (C) From steps (A) and (B) it
follows that $M=M(\langle \sigma _{i}^{a}\sigma _{j}^{b}\rangle )$. However,
by using the HK theorem
%% RESET:
for non-degenerate ground states,
any expectation value can be taken as a function of
$\{a_{l}\}$, since the wave function itself is a function of $\{a_{l}\}$
(see, e.g., Ref.~\cite{Schonhammer:95}). Therefore, $M=M(\{a_{l}\})$, as
required. {\ \rule{0.5em}{0.5em}}

In Ref.~\cite{WuSarandyLidar:04}, relations similar to Eq.~(\ref{eq02}),
which connects entanglement and 1QPTs, were established at the critical
point for several examples of multi-particle systems, up to two-body
interactions. In DFT, Eq.~(\ref{eq02}) holds for arbitrary systems, and not
only close to the critical point. 
%%%%%%%%%%%%%%%%
%% Adjusted in v6b:
%%DELETE It is important to emphasize that, in the case of 
%%degeneracy, some special care is required. 
%%ACCEPT CHANGES HERE
While the HK theorem is
also applicable to degenerate ground states~\cite{Levy:79,Kohn:85}, not all
linear combinations of densities corresponding to degenerate ground states are
permissible when implementing the variational 
principle~\cite{Lieb:83}.
%%END CHANGE HERE, CONTINUE BELOW
Note also that systems described by either Fermi
or Pauli operators can be considered using DFT. Indeed, the treatment of
both cases can be unified by the Jordan-Wigner transformation~\cite%
{Jordan:28}, with $H$, $H_{0}$, and $H_{\mathrm{ext}}$ expressed in terms of
linear combinations of generators of $SU(2^{N})$, where $N$ denotes the
number of sites in the case of spins in a lattice, or the number of single
modes for Jordan-Wigner
fermions.

Moreover, the HK theorem implies that one can split up the Hamiltonian
(\ref{eq00}) in different ways. For example, a new $H_{0}$ might include
part of the sum $\sum \lambda _{l}\widehat{A}_{l}.$ In our discussion, it is
often convenient to focus on one of the external operators by moving the
others into $H_{0}$.

For 2QPTs, we should examine the derivatives of $M$. For simplicity of
exposition, we regard one of the parameters $\lambda _{l}$ as an independent
variable, which we denote by $\lambda $, and consider all the others as part
of $H_{0}$. Therefore, $M$ can be seen as an exclusive function of $\lambda $%
, yielding via Eq.~(\ref{eq01})
\begin{equation}
\frac{\partial M}{\partial \lambda }=\frac{\partial M}{\partial a}\frac{%
\partial a}{\partial \lambda }=\frac{\partial M}{\partial a}\frac{\partial
^{2}E}{\partial \lambda ^{2}}.  \label{eq03}
\end{equation}%
%%ACCEPT ADDITION
Notice that this equation holds only for non-degenerate ground states, since
for the case of degeneracy, although the density $a$ still uniquely 
specifies the
potential $\lambda$, the potential $\lambda$ does not uniquely 
specify the density $a$
anymore. Therefore, in the degenerate case, $a$ cannot be taken as a 
function of $\lambda$,
which implies that the chain rule used to take the derivative in 
Eq.~(\ref{eq03}) is not valid.
However, as long we restrict ourselves to non-degenerate states (as 
is the usual case for large finite
systems tending to criticality), or approach the (critical) 
degeneracy point from below or above, this
problem can be avoided. In the case where the degeneracy is symmetry 
driven, we could also circumvent this
problem by observing that degenerate states can be split by a 
symmetry breaking term which is then allowed
to tend to zero in the study of QPTs.

Eq.~(\ref{eq03}) shows that an entanglement measure is proportional to the
second derivative of energy as long as ${\partial M}/{\partial a}\neq 0$. By
using appropriate bipartite entanglement measures, 2QPTs have usually been
identified so far through either non-analytic or vanishing values of ${%
\partial M}/{\partial \lambda }$ at the critical point. Both cases are
contained in Eq.~(\ref{eq03}).

It should be emphasized that Eqs.~(\ref{eq02}) and~(\ref{eq03})
hold for any system described by the Hamiltonian (\ref{eq00}) as
long as DFT is valid,
in the degeneracy sense discussed above.
Around the critical points, the left and
right limits of the two equations still hold even if the DFT is
questionable at the critical point.
\emph{Eqs.~(\ref{eq02}) and~(\ref{eq03}) can be seen
as the basic equations for the relation between QPTs and entanglement}.

\section{Example I: One-body external couplings}
As a first example of the applicability of Eqs.~(\ref{eq02}) and (\ref{eq03}),
let us consider $H_{\mathrm{ext}}=\sum_{i}\overrightarrow{\lambda 
}_{i}\cdot \overrightarrow{%
\sigma }_{i}$, which represents a system of qubits acted upon via
independent single-qubit control terms. According to DFT, the energy is a
functional of matrix elements of one-spin reduced density matrices, i.e., $%
E=E(\{\overrightarrow{\rho }_{i}\})$, where $\overrightarrow{\rho }%
_{i}=\left\langle \psi \right\vert \overrightarrow{\sigma }_{i}\left\vert
\psi \right\rangle =\overrightarrow{\bigtriangledown }_{i}E$ is the Bloch
vector, with components $\rho _{i}^{\alpha }=\left\langle \psi \right\vert
\sigma _{i}^{\alpha }\left\vert \psi \right\rangle $. We consider a
bipartition of the system, splitting it up into two parts. Then, assuming
that the system is in a pure state, we can use the linear entropy as a
measure of block entanglement, which reads $L^{(d)}=[d/(d-1)](1-{\textrm{Tr}}%
\,\rho ^{2})$, where $0\leq L^{(d)}\leq 1$ and $\rho $ denotes a $d\times d$%
-dimensional density matrix. Explicit computation of the block entanglement
of one qubit (the $i$th) with the rest of system yields $L_{i}^{(2)}=1-\left%
\vert \overrightarrow{\rho }_{i}\right\vert ^{2}=1-\overrightarrow{%
\bigtriangledown }_{i}E\cdot \overrightarrow{\bigtriangledown }_{i}E$, which
is a function of the parameters $\overrightarrow{\lambda }$. In the case of
fermions, we replace the Pauli matrices by fermionic operators 
according to the Jordan-Wigner transformation. Then, $%
L_{i}^{(2)}=1-(\partial E/\partial \lambda _{zi})^{2}$ (number conservation
law for fermions implies the vanishing of $\partial E/\partial \lambda _{xi}$
and $\partial E/\partial \lambda _{yi}$).

\begin{figure}[th]
\centering {\includegraphics[angle=0,scale=0.34]{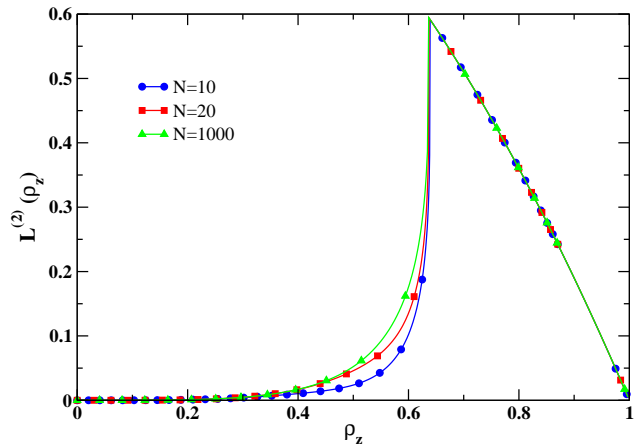}}
\caption{Block entanglement $L^{(2)}$ as function of $\protect\rho _{z}$ for
the transverse field Ising model. The maximum occurs at the quantum critical
point, where $\protect\rho _{z}\approx 0.6366$.}
\label{f1}
\end{figure}

\begin{figure}[th]
\centering {\includegraphics[angle=0,scale=0.34]{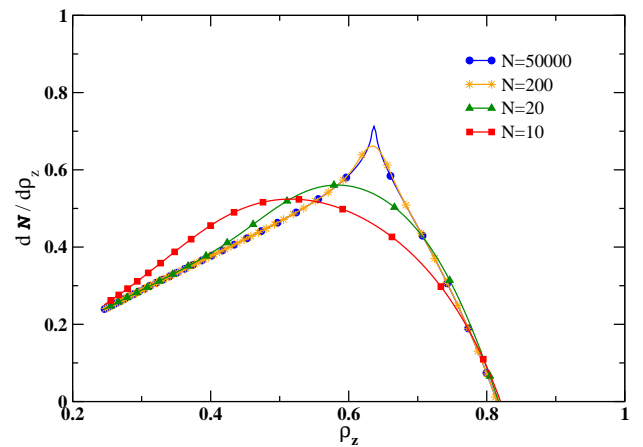}}
\caption{Derivative of the negativity with respect to $\protect\rho _{z}$
for the transverse field Ising model. The maximum approaches the quantum
critical point, where $\protect\rho _{z}\approx 0.6366$.}
\label{f2}
\end{figure}

In the case of a 1QPT, characterized by a discontinuity in $\overrightarrow{%
\bigtriangledown }_{i}E$, we have a corresponding discontinuity in $%
L_{i}^{(2)}$ unless $\overrightarrow{\bigtriangledown }_{i}E\cdot
\overrightarrow{\bigtriangledown }_{i}E$ is continuous. Therefore, in this
case, when all $\overrightarrow{\lambda }_{i}$ are taken as independent
external parameters, the entanglement measure $L_{i}^{(2)}$ is an analytic
function of the first derivatives of the energy, yielding a natural
relationship between 1QPTs and $L_{i}^{(2)}$. A general discussion of 2QPTs
is, on the other hand, not as straigthforward, since the structure of the
derivatives of $L_{i}^{(2)}$ will depend on the details of the model. Thus,
it turns out to be more useful to analyze a concrete example. Let us
consider the transverse field Ising chain, where $H=-\sum_{i=1}^{N}(\sigma
_{i}^{x}\sigma _{i+1}^{x}+\lambda \sigma _{i}^{z})$, with $N$ denoting the
number of spins along the chain and with cyclic boundary conditions assumed.
%%%%%%%%%%%%%%%%%%%%%%%%%%%%%%%%%%%%%%%%%%%%%%%%
%% Added v6
In this model, a discussion of entanglement as a function of the coupling $\lambda$ 
was first presented in Refs.~\cite{Osterloh:02,Osborne:02}. 
%%%%%%%%%%%%%%%%%%%%%%%%%%%%%%%%%%%%%%%%%%%%%%%%
Due to translational symmetry we have $\rho _{z}=\left\langle \psi
\right\vert \sigma ^{z}\left\vert \psi \right\rangle =\frac{\partial
\varepsilon }{\partial \lambda }$, where $\varepsilon =E/N$. Therefore $%
\frac{\partial L^{(2)}}{\partial \lambda }=\frac{\partial L^{(2)}}{\partial
\rho _{z}}\frac{\partial ^{2}\varepsilon }{\partial \lambda ^{2}}$.
Divergence of ${\partial ^{2}\varepsilon }/{\partial \lambda ^{2}}$ at the
quantum critical point $\lambda =1$ will thus result in that of ${\partial
L^{(2)}}/{\partial \lambda }$ unless ${\partial L^{(2)}}/{\partial \rho
_{z}=0}$, which is not the case in this example. This is demonstrated in
Fig.~\ref{f1}, where we plot $L^{(2)}$ as a function of $\rho _{z}$. Both
the maximum and the singularity of the derivative occur at the critical
point. We can also apply the DFT approach to pairwise entanglement measures.
For instance, let us consider entanglement between nearest-neighbor pairs in
the transverse field Ising model as measured by the negativity~$\mathcal{N}$
\cite{Vidal:02a}. From Eq.~(\ref{eq03}) we have $\frac{\partial \mathcal{N}}{%
\partial \lambda }=\frac{\partial \mathcal{N}}{\partial \rho _{z}}\frac{%
\partial ^{2}\varepsilon }{\partial \lambda ^{2}}$. Notice that the
divergence in ${\partial ^{2}\varepsilon }/{\partial \lambda ^{2}}$ at the
critical point naturally leads to a divergence in ${\partial \mathcal{N}}/{%
\partial \lambda }$, since ${\partial \mathcal{N}}/{\partial \rho _{z}}$ is
a non-vanishing function at the QPT, as shown in Fig.~\ref{f2}. In fact, the
maximum of ${\partial \mathcal{N}}/{\partial \rho _{z}}$ approaches the
critical point as the number of sites increases.

\section{Example II: two-body external couplings}
In the case of two-body external couplings, we take
$H_{\mathrm{ext}}=\sum_{ij}\lambda _{ij}^{\alpha }\sigma _{i}^{\alpha
}\sigma _{j}^{\alpha }$, where $\alpha \in \{x,y,z\}$. This Hamiltonian
represents a system of qubits controlled externally via two-body
interactions. Entanglement between qubits $i,j$ and the rest of the system
can then be computed by taking the linear entropy $L^{(d)}$ for $d=4$. This
yields $L^{(4)}=1-\frac{1}{3}\sum (\frac{\partial E}{\partial \lambda
_{ij}^{\alpha }})^{2}$. We now analyze the behavior of $L^{(4)}$ in some
important models exhibiting QPTs. For example, for the XXZ spin chain, we
have $H=(-1/2)\sum_{i}^{N}(\sigma _{i}^{x}\sigma _{i+1}^{x}+\sigma
_{i}^{y}\sigma _{i+1}^{y}+\Delta \sigma _{i}^{z}\sigma _{i+1}^{z})$, where
cyclic boundary conditions are assumed. The external Hamiltonian is taken as
$H_{\mathrm{ext}}=-(\Delta /2)\sum_{i}\sigma _{i}^{z}\sigma _{i+1}^{z}$.
Direct evaluation of $L^{(4)}$ then yields (See Appendix~\ref{a2})
\begin{equation}
L^{(4)}=1-\frac{4}{3}\left[ \left( 1+\frac{\Delta ^{2}}{2}\right) \left(
\frac{\partial \varepsilon }{\partial \Delta }\right) ^{2}+\frac{\varepsilon
^{2}}{2}-\varepsilon \Delta \frac{\partial \varepsilon }{\partial \Delta }%
\right] ,  \label{l4xxz}
\end{equation}%
where $\varepsilon =E/N$. Notice that $L^{(4)}$ is a function of the DFT
variable $a=\langle \sigma _{i}^{z}\sigma _{i+1}^{z}\rangle =-2(\partial
\varepsilon /\partial \Delta )$ since, due to the HK theorem, the energy
density $\varepsilon $ can be taken as a function of $a$. Thus,
discontinuities in $(\partial \varepsilon /\partial \Delta )$ will be
directly reflected in $L^{(4)}$. This model exhibits two distinct QPTs,
which occur at $\Delta =1$ and $\Delta =-1$. In order to evaluate $L^{(4)}$,
we consider the ground state wave-function with vanishing magnetization,
which favors the presence of entanglement in the system. A 1QPT occurs at $%
\Delta =1$, which separates a ferromagnetic phase from a gapless
quasi-long-range-ordered phase. At this ferromagnetic critical point, the
energy density as $N\rightarrow \infty $ is continuous, and is given by
$\varepsilon (\Delta =1)=-1/2$~\cite{Yang:66a,Yang:66b}.
However, its first derivative
is discontinuous, with
$(\partial \varepsilon /\partial \Delta )_{\Delta\rightarrow 
1^{+}}\rightarrow -1/2$
and $(\partial \varepsilon /\partial\Delta )_{\Delta \rightarrow 
1^{-}}\rightarrow 0$.
 From Eq.~(\ref{l4xxz}), we can see that this discontinuity is 
immediately manifested in the
entanglement measure, since $L^{(4)}$ jumps from
$2/3$ to $5/6$
at $\Delta =1$. A continuous QPT in the XXZ chain occurs at $\Delta =-1$,
separating the gapless quasi-long-range-ordered phase from the
antiferromagnetic phase. For this case, it is useful to compute the first
derivative of $L^{(4)}$ with respect to $\Delta $, which yields $({\partial
L^{(4)}}/{\partial \Delta })=({\partial L^{(4)}}/{\partial a})({\partial
^{2}\varepsilon }/{\partial \Delta ^{2}})$, with $\frac{\partial L^{(4)}}{%
\partial a}=\frac{4}{3}\left[ \varepsilon \Delta -2\left( 1+\frac{\Delta ^{2}%
}{2}\right) \frac{\partial \varepsilon }{\partial \Delta }\right] $. The QPT
in this case is not directly signalled by $({\partial ^{2}\varepsilon }/{%
\partial \Delta ^{2}})$, which is analytic at $\Delta 
=-1$~\cite{Yang:66a,Yang:66b}.
However, entanglement detects this transition as an extremum at the critical
point~\cite{Gu:03,Chen:04,Yang:05}. This behavior is also reflected in terms
of the DFT variable $a$. We have
$\varepsilon (\Delta =-1)= 2(\ln{2}-1/4)$~\cite{Yang:66a,Yang:66b},
and find for the first derivative of the energy
$(\partial\varepsilon /\partial \Delta )_{\Delta \rightarrow 
-1}\approx 0.2954$.
Therefore, we obtain $({\partial L^{(4)}}/{\partial a})=({\partial L^{(4)}}/{%
\partial \Delta })=0$.

We now analyze the behavior of $L^{(4)}$ in a Fermi system. An interesting
example is then the one-dimensional Hubbard model, $H_{\mathrm{ext}%
}=U\sum_{\alpha }n_{\alpha \uparrow }n_{\alpha \downarrow }$, where $%
n_{\alpha \downarrow }~(n_{\alpha \uparrow })$ is the spin down (up)
electronic number at site $\alpha $. 
%%%%%%%%%%%%%%%%%%%%%%%%%%%%%%%%%%%%%%%%%%%%%%%%%%%%%%%%%%%%%%%%%%%%%%
%%% Added v6:
The Hubbard model describes a metal-insulating transition, which has been 
considered from the point of view of entanglement in Refs.~\cite{Gu:04,Larsson:05}.
%%%%%%%%%%%%%%%%%%%%%%%%%%%%%%%%%%%%%%%%%%%%%%%%%%%%%%%%%%%%%%%%%%%%%%
We can rearrange the indices for the
modes $\alpha \uparrow $ and $\alpha \downarrow $ into nearest neighbor
indices $i$ and $i+1$, respectively, in a linear lattice. Therefore, the
Hamiltonian can be written as $H_{\mathrm{ext}}=U\sum_{i}n_{2i-1}n_{2i}$
where only the pairs of sites (1,2), (3,4), etc., interact with each other.
We can then compute $L^{(4)}$ between an interacting pair $(i,i+1)$ and the
rest of the system (see also Refs.~\cite{Gu:04,Larsson:05}). At
half-filling, $L^{(4)}=\frac{2}{3}(1+4a-8a^{2})$ (for any $i$) (See Appendix~\ref{a3}).
Then $\frac{\partial L^{(4)}}{\partial a}=\frac{2}{3}(4-16a)$. By 
using Eq.~(\ref{eq03}%
), we obtain $\frac{\partial L^{(4)}}{\partial U}=\frac{\partial L^{(4)}}{%
\partial a}\frac{\partial ^{2}\varepsilon }{\partial U^{2}}$. At the
critical point $U=0$, which separates an insulating phase from a metallic
phase, the first derivative of $L^{(4)}$ with respect to $U$ is ${\partial
L^{(4)}}/{\partial U}=0$~\cite{Gu:04}. In terms of the new variable $a$, we
can show that the QPT in the Hubbard model is also identified via an
extremum of $L^{(4)}$. Indeed, for $U=0$, we have $a=1/4$~\cite{Economou:79}
which then implies $\partial _{a}L^{(4)}=0$.

\section{The Lipkin model: a Hartree-Fock approach to entanglement}
Most realistic physical many-body problems cannot be solved analytically.
Linear approximations, such as Hartree-Fock-Bogoliubov theory and the random
phase approximation, are often practical and effective ways to treat these
systems, since these procedures change an intractable $2^{N}$-dimensional
problem to a tractable $N^{2}$-dimensional one. In this case, it is
appealing to introduce new and simple quantities, e.g., $L^{(2)}$ and $%
L^{(4)}$, as measures characterizing the quantum information content of
these known approximate wave functions.
We expect these quantities to become as important as, e.g., binding
energies, when quantum information becomes readily accessible to
experiments.

As an example, we consider the Lipkin model --
important, e.g., in nuclear physics --
whose Hamiltonian reads $H=\lambda
S_{z}-\frac{1}{N}(S_{x}^{2}-S_{y}^{2})$, where $S_{z}=\sum_{m=1}^{N}\frac{1}{%
2}(c_{+m}^{\dagger }c_{+m}-c_{-m}^{\dagger }c_{-m})$ and $%
S_{x}+iS_{y}=\sum_{m=1}^{N}c_{+m}^{\dagger }c_{-m}$~\cite{Ring:book} 
%%%%%%%%%%%%%%%%%%%%%%%%%%%%%%%%%%%%%%%%%%%%%
%% Added v6
(for a discussion of entanglement in the Lipkin model, 
see also Ref.~\cite{Vidal:04}).
%%%%%%%%%%%%%%%%%%%%%%%%%%%%%%%%%%%%%%%%%%%%%%
This
Hamiltonian describes a two-level Fermi system $\{|+\rangle ,|-\rangle \}$,
each level having degeneracy $N$. The operators $c_{+m}^{\dagger }$ and $%
c_{-m}^{\dagger }$ create a particle in the upper and lower levels,
respectively. Alternatively, the Hamiltonian may be viewed as a
one-dimensional ring of two-level atoms with infinite range interaction
between pairs. The factor $1/N$ in the interaction term keeps the scaling of
both terms in $H$ linear in $N$. The phase transition studied is in the
limit of $N\rightarrow \infty $.
The Lipkin model is exactly solvable (see, e.g., Ref.~\cite{Ortiz:05}).
The Hartree-Fock (HF) ground state, which is exact for this model as $N$
tends to infinity, is given by $|HF\rangle =\prod_{m=1}^{N}a_{0m}^{\dagger
}|-\rangle $, where $a_{0m}^{\dagger }$ is defined by the following change
of variables: $c_{+m}^{\dagger }=\sin {\alpha }\,a_{0m}^{\dagger }+\cos {%
\alpha }\,a_{1m}^{\dagger }$ and $c_{-m}^{\dagger }=\cos {\alpha }%
\,a_{0m}^{\dagger }-\sin {\alpha }\,a_{1m}^{\dagger }$. The variational
parameter $\alpha $ which yields the minimum energy is given by $\cos {%
2\alpha }=\lambda $ when $\lambda <1$ and $\alpha =0$ when $\lambda \geq 1$.
We define the DFT variable $a=\partial \varepsilon /\partial \lambda $, with
$\varepsilon =E/N$ denoting the energy per particle. For the HF ground
state, we then obtain $a=-\frac{\lambda }{2}$ for $\lambda <1$ and $a=-\frac{%
1}{2}$ for $\lambda \geq 1$ It is easy to show that $\partial
^{2}\varepsilon /\partial \lambda ^{2}$ is discontinuous at $\lambda =1$,
which corresponds to $a=-1/2$ in terms of the DFT variable. Let us analyze
whether this discontinuity is reflected in the derivatives of the
entanglement measures, as given by Eq.~(\ref{eq03}). For $4$-dimensional
block entanglement, it is convenient to consider the entanglement between a
block composed of two general modes ($+m$,$-n$) and the rest of the system,
which yields $L_{+m,-n}^{(4)}=(2/3)(1-4a^{2})(1-\delta _{m,n})$, where $%
\delta _{m,n}$ is the Kronecker symbol. Therefore, the block ($+m,-n$) is
entangled with the rest of system only if $m\neq n$. Taking the derivative,
we obtain $(\partial L_{+m,-n}^{(4)}/\partial a)_{a=-1/2}=8/3$ $(m\neq n)$.
Therefore, from Eq.~(\ref{eq03}), the non-analyticity of $\partial
^{2}\varepsilon /\partial \lambda ^{2}$ at the critical point will be
associated to a non-analyticity in $\partial L_{+m,-n}^{(4)}/\partial
\lambda $ ($m\neq n$). A similar result follows in the case of $2$%
-dimensional block entanglement, where we have $L^{(2)}=1-4a^{2}$ for a
general mode $+m$ (or $-m$) with the rest of the system. Pairwise
entanglement between general modes $+m$ and $-n$ as measured by the
negativity is found to be $\mathcal{N}_{+m,-n}=\sqrt{1-4a^{2}}\delta _{m,n}$%
. Notice that this is in contrast with block entanglement, where modes $+m$
and $-n$ only are entangled for $m\neq n$. This difference is due to the
structure of the HF ground state, which implies that the modes $+m$ and $-n$
interact only for $m=n$. Therefore, bipartite entanglement in the system
appears only when $+m$ and $-m$ are in different parts. Evaluating now the
derivative of the negativity we obtain $(\partial \mathcal{N}%
_{+m,-n}/\partial a)_{a\rightarrow -1/2}\rightarrow \infty $. Thus, $%
\partial \mathcal{N}_{+m,-n}/\partial \lambda $ \ is non-analytic at the
critical point.

\section{Conclusion}
We have shown in general and illustrated in a number
of models that DFT provides a natural link between entanglement and QPTs.
Since experimental data are taken at finite temperature, it is important to
be able to delineate the temperature fluctuation around a classical critical
point versus the quantum fluctuations around a QPT. The exploration of
finite-temperature DFT~\cite{Mermin} for the connection between phase
transitions and quantum information appears to be a promising direction for
future study.

\section*{Acknowledgements}
We gratefully acknowledge financial support from FAPESP (to
M.S.S.), the Sloan Foundation (to D.A.L.), and NSF DMR 0403465 (to L.J.S.).
M.S.S. also thanks Prof. F. C. Alcaraz and Prof. K. Capelle for their
comments.

%%%%%%%%%%%%% 
%Appendices
%%%%%%%%%%%%%

\appendix

%%%%%%%%%%%%%%%%%%%%%%%
\section{}
\label{a1}
We provide here a proof of the HK theorem in a lattice,
based on the variational method (for a proof based on the constrained-search
technique~\cite{Levy:79}, see Ref.~\cite{Schonhammer:95}). Let us consider
two sets of parameters $\{\lambda_l\}$ and $\{\lambda_l^\prime\}$, which
define two Hamiltonians as follows:
\begin{equation}
H = H_0+\sum_l \lambda_l A_l\,\, ,\,\,\,\,\,\,\, H^\prime = 
H_0+\sum_l \lambda_l^\prime A_l.
\label{HH}
\end{equation}
The ground states of $H$ and $H^\prime$ will be denoted by 
$|\psi\rangle$ and $|\psi^\prime\rangle$,
respectively, which are taken as non-degenerate, even though the 
proof can be extended for degenerate
ground states~\cite{Levy:79,Kohn:85}. We also assume here that, for 
different sets of parameters,
$\{\lambda_l\} \ne \{\lambda_l^\prime\}$, we have independent ground states
$|\psi\rangle \ne \alpha |\psi^\prime\rangle$
($\alpha=$ constant). This is indeed the usual behavior of quantum 
systems around criticality, where
the ground state varies continuously as we vary the control 
parameters. By applying the variational
principle for the Hamiltonian $H^\prime$, we obtain
\begin{equation}
\langle \psi^\prime| H^\prime |\psi^\prime \rangle <  \langle \psi| 
H^\prime |\psi \rangle
= \langle \psi| \left(H+\sum_l\left(\lambda_l^\prime - 
\lambda_l\right) A_l\right)  |\psi \rangle
\label{vp}
\end{equation}
Therefore, Eq.~(\ref{vp}) yields
\begin{equation}
E_0^\prime < E_0 + \sum_l \left(\lambda_l^\prime - \lambda_l\right) a_l,
\label{evp1}
\end{equation}
where $E_0^\prime$ and $E_0$ are the ground state energies of 
$H^\prime$ and $H$, respectively,
and $a_l=\langle \psi|A_l|\psi\rangle$. Analogously, by applying the 
variational principle for $H$,
we obtain
\begin{equation}
E_0 < E_0^\prime + \sum_l \left(\lambda_l - \lambda_l^\prime\right) a_l^\prime,
\label{evp2}
\end{equation}
with $a_l^\prime=\langle \psi^\prime|A_l|\psi^\prime\rangle$. From
Eqs.~(\ref{evp1}) and~(\ref{evp2}) we have
\begin{equation}
0 < \sum_l \left(\lambda_l^\prime - \lambda_l\right) \left(a_l - 
a_l^\prime\right)
\label{fcon}
\end{equation}
Hence, if the sets of parameters $\{\lambda_l\}$ and 
$\{\lambda_l^\prime\}$ are different from
each other, then we cannot have identical sets $\{a_l\}$ and 
$\{a_l^\prime\}$. Therefore, the
density $\{a_l\}$ uniquely specifies the potential $\{\lambda_l\}$ 
and can then be used as the
basic variable to describe the properties of the system.

%%%%%%%%%%%%%%%%%%%%%%%%%%
\section{}
\label{a2}
We provide here the basic details of the 
evaluation of the linear entropy for the
XXZ model. The density matrix for a pair of nearest-neighbor sites in 
the ground state of the XXZ chain
can be written as
\begin{equation}
\mathcal{\rho}=\left(
\begin{array}{cccc}
A & 0 & 0 & 0 \\
0 & B & C & 0 \\
0 & C & B & 0 \\
0 & 0 & 0 & D
\end{array}%
\right) ,  \label{rhoXXZ}
\end{equation}%
where, from the XXZ Hamiltonian, we obtain
\begin{eqnarray}
A&=&D=\frac{1}{4}\left(1-2\frac{\partial \varepsilon}{\partial 
\Delta}\right), \,\,\,\,\,
B=\frac{1}{4}\left(1+2\frac{\partial \varepsilon}{\partial 
\Delta}\right), \nonumber \\
C&=&-\frac{1}{2}\left(\varepsilon-\Delta\frac{\partial \varepsilon}{\partial 
\Delta}\right).
\label{matelemXXZ}
\end{eqnarray}
Eqs.~(\ref{matelemXXZ}) allows for a direct calculation of the linear entropy
$L^{(4)}=(4/3)(1-{\textrm{Tr}}\rho^2)$, yielding the result displayed 
in Eq.~(\ref{l4xxz}).

%%%%%%%%%%%%%%%%%%%%
\section{}
\label{a3}
We provide here the basic details for the 
evaluation of the linear entropy in the
Hubbard model. Translation invariance and simultaneous conservation 
of particle number
$N=\sum_j\left( n_{j\uparrow}+n_{j\downarrow}\right)$ and 
$z$-component of total spin
$S^z=\sum_j\left( n_{j\uparrow}-n_{j\downarrow}\right)$ imply that 
the density operator
for any single site can be represented by a $4\times 4$ diagonal 
matrix, whose eigenvalues are given by
\begin{eqnarray}
w=\langle n_{\alpha \uparrow}n_{\alpha \downarrow} \rangle = 
\frac{\partial \varepsilon}{\partial U} \equiv a, \,\,\,\,
u^{+}=\langle n_{\alpha \uparrow}\rangle - w, \nonumber \\ 
u^{-}=\langle n_{\alpha \downarrow}\rangle - w, \,\,\,\,
z=1-u^{+}-u^{-}-w.
\label{matelemhub}
\end{eqnarray}
At half-filling, we have $ \langle n_{\alpha \uparrow} \rangle = 
\langle n_{\alpha \downarrow} \rangle =1/2$.
Therefore, in this regime, all the eigenvalues can be expressed in 
terms of the density
$a=\langle n_{\alpha \uparrow}n_{\alpha \downarrow} \rangle$. Then, 
the evaluation of the linear entropy
$L^{(4)}(a)$ follows straightforwardly.

\end{document}